\documentclass[a4paper,12pt,fleqn]{article}
\usepackage{amsmath,graphicx}
\newcommand{\pid}{\par\indent}
\newcommand{\pni}{\par\noindent}

\begin{document}
\title{Kerr-AdS and Kerr-dS solutions revisited}
\author{ A. G. Agnese\footnote{Email: agnese@ge.infn.it},
M. La Camera\footnote{Email: lacamera@ge.infn.it}} 
\date{}
\maketitle 
\begin{center}
\emph {Dipartimento di Fisica dell'Universit\`a di 
Genova\\Istituto Nazionale di Fisica Nucleare,Sezione di 
Genova\\Via Dodecaneso 33, 16146 Genova, Italy}\\
\end{center} 
\bigskip
\begin{abstract}\pni
We reconsider the Kerr metric with cosmological term $\Lambda$ 
imposing the condition that the angular velocity $\omega$ of 
the dragging of inertial frames vanishes at spatial boundaries. 
Some properties of the extreme black holes in the revisited 
solutions are discussed. 
\end{abstract} 
\bigskip\bigskip\bigskip \pni
\vspace{1in}\pni
\newpage
\baselineskip = 2\baselineskip

With the advent of Maldacena's conjecture of Anti de Sitter -
Conformal Field Theory correspondence (AdS-CFT) [1], there has 
been a great deal of interest in studying the properties of black
holes in AdS space [2-7], with special emphasis on the Kerr-AdS 
or Kerr-Newman-AdS solutions [8,9,10]. Rotating black holes in 
four dimensions with asymptotic AdS behavior were first 
constructed by Carter many years ago [11]. \pid
The purpose of this letter is to discuss various properties,  
which have not been considered in the literature before, of one 
of Carter's families of Kerr vacuum solutions with cosmological 
term $\Lambda$. We refer to the stationary and axisymmetric 
metric (family $[A]$ of Ref.11) which can be written as
\begin{multline}
ds^2 = \dfrac{\Delta_r}{\rho^2}\, \left[d\chi -\frac{a}{\Xi} 
\sin^2\vartheta d\psi \right]^2 -\dfrac{\rho^2}{\Delta_r}\, dr^2 
\\ - \dfrac{\rho^2}{\Delta_\vartheta}\, d\vartheta^2 
-\dfrac{\sin^2\vartheta \Delta_\vartheta}{\rho^2}\, \left[a d\chi
- \dfrac{(r^2+a^2)}{\Xi} d\psi \right]^2
\end{multline}
where
\begin{equation}\begin{split}
\rho^2 &= r^2 + a^2 \cos^2\vartheta \\
\Delta_r &= - \dfrac{\Lambda}{3} r^4 + (1-\dfrac{a^2\Lambda}{3}) 
r^2 - 2Mr +a^2 \\
\Delta_\vartheta &= 1 + \dfrac{a^2 \Lambda}{3} \cos^2\vartheta \\
\Xi &= 1 + \dfrac{a^2 \Lambda}{3}
\end{split}\end{equation}
\newpage
The parameter $M$ is related to the mass, $a$ to the angular 
momentum per unit mass while $\chi$ and $\psi$ are two ignorable 
coordinates.
To  express $\chi$ and $\psi$ by means of the usual time 
and azimuthal angle coordinates $t$ and $\varphi$, we use the 
coordinate transformations
\begin{align}
\chi &= \alpha \, t \\
\psi &= \beta \, \varphi + \gamma \, t
\end{align}
where the constants $\alpha ,\beta$ and $\gamma$ are to be 
determined with the conditions that the angular velocity $\omega$
of the dragging of inertial frames must vanish when $r$ reaches 
infinity if $\Lambda < 0$ and when $r$ reaches the cosmological 
horizon if $\Lambda > 0$ ; moreover $t$ and $\varphi$ will be 
properly normalized.\pid 
We first consider the Kerr-AdS case ($\Lambda < 0$).\pid 
The required transformations are
 \begin{align}
\chi &= \left( 1+\dfrac{a^2 \Lambda}{3}\right)^2 t \\
\psi &= \sqrt{ 1+\dfrac{a^2 \Lambda}{3}}\, \varphi + \dfrac{a 
\Lambda}{3} \left( 1+\dfrac{a^2 \Lambda}{3}\right) t
\end{align}
and the corresponding line element (1) in Boyer-Lindquist 
coordinates \linebreak becomes \pagebreak 
\begin{multline}
\hspace{-0.4cm} ds^2 = \dfrac{\left(1+\dfrac{a^2 \Lambda}{3} 
\cos^2 \vartheta \right) \left[ \left(1+\dfrac{a^2 \Lambda}{3} 
\cos^2 \vartheta \right) \Delta_r -a^2 \sin^2 \vartheta \left( 
1 -\, \dfrac{\Lambda r^2}{3}\right)^2\right]}{r^2+a^2 \cos^2 
\vartheta}\, dt^2 \\ {} \\ \hspace{-1.2cm}
-\, \dfrac{r^2+a^2 \cos^2 \vartheta}{\Delta_r}\, dr^2 -\, \frac
{r^2+a^2 \cos^2 \vartheta}{1+\dfrac{a^2 \Lambda}{3} \cos^2 
\vartheta}\, d\vartheta ^2 + \dfrac{4 M r a \sin^2 \vartheta 
\left( 1+\dfrac{a^2 \Lambda}{3} \cos^2 \vartheta 
\right)}{\sqrt{1+ \dfrac{a^2 \Lambda}{3}}\,(r^2+a^2 \cos^2 
\vartheta)}\, dtd\varphi \\ {}\\ 
-\, \dfrac{\sin^2 \vartheta \left[\left(1+\dfrac{a^2 \Lambda}{3} 
\cos^2 \vartheta \right)(r^2+a^2)^2 - a^2 \sin^2 \vartheta 
\Delta_r \right] }{ \left( 1+\dfrac{a^2 
\Lambda}{3}\right)(r^2+a^2 \cos^2 \vartheta)}\, d\varphi ^2 
\end{multline}
The solution is valid for $1+\dfrac{a^2 \Lambda}{3} > 0 $ and 
becomes singular when the latter quantity is zero.
The event horizon is located at $r=r_+$, the larger of the two 
positive roots $r_+$ and $r_-$  of the polynomial $\Delta_r$.\pid
In this letter we limit ourselves to consider some properties of 
extreme black holes.\pid In the parameter plane \; $(M^2 \Lambda 
/3, a^2/M^2)$\; the curve $r_+ = r_-$ represents the locus of the
extreme black holes, i.e. the borderline between black holes and 
naked singularities. The equation of this curve is obtained  
requiring that $\Delta_r = \Delta_r '= 0$, where a prime denotes 
derivative with respect to $r$ and positive roots are to be 
considered. Putting for simplicity $x=M^2\Lambda /3, \; 
y=a^2/M^2$, one obtains the following equation 
\begin{multline} x^3y^3 + 9x^2y^ 2 - 9xy + 27 x - 1 -
9x\,\sqrt{8y(xy - 1) + 9} \\ +(xy - 1)^2\,\sqrt{xy(xy - 14)+1} = 
0 \end{multline}
The corresponding plot is  given in Fig.1 and  
is comprised between the ``critical'' values $(-\, 64/27,\, 
27/64)$ which correspond to $1+\dfrac{a^2 \Lambda}{3} = 0 $, and 
the values $(0,1)$.\pid The angular velocity $\omega$ is given by
\begin{equation}
\omega = \dfrac{2 M r a\, \sqrt{1+\dfrac{a^2 \Lambda}{3}}
\left(1+\dfrac{a^2 \Lambda}{3} \cos^2 \vartheta \right)}{
\left(1+\dfrac{a^2 \Lambda}{3} \cos^2 \vartheta \right) 
(r^2+a^2)^2-\, a^2 \sin^2 \vartheta \Delta_r}
\end{equation}
One can immediately see that the angular velocity vanishes not 
only asymptotically, but also at the critical point above 
defined. A plot of $\omega$ as a function of the radius $r_+$ of 
the extreme black hole: 
\begin{equation}
\omega = \dfrac{ a\, \sqrt{1+\dfrac{a^2 \Lambda}{3}}\left(1-\, 
\dfrac{\Lambda r_+^2}{3}\right)}{ r_+^2+a^2}
\end{equation} 
is given in Fig.2, in terms of the dimensionless 
quantities $\omega ' = M \omega$ and $r_+ ' = r_+/M$. \pid 
The area of the horizon is
\begin{equation}
A = 4\pi \dfrac{r_+^2 + a^2}{\sqrt{1+\dfrac{a^2 \Lambda}{3}}}
\end{equation} 
It diverges at the critical point, gets its minimum near 
$r_+ = M/2$ then increases till the value $8\pi M^2$ reached at 
$r_+=M$. A plot of $A$ as a function of $r_+$ is given in 
Fig.3, where are used the dimensionless quantities $A'~=~A/(8\pi 
M^2)$ and $r_+ ' = r_+/M$. We notice that the minimum value of
the area, which corresponds to the minimum value of the entropy, 
is also in correspondence with the maximum value of the angular 
velocity. \pid
In a similar fashion we can now treat the Kerr-dS case ($\Lambda 
> 0$).\pid 
The coordinate transformations are 
\begin{align}
\chi &= \left( 1+\dfrac{a^2 \Lambda}{3}\right) t \\
\psi &= \sqrt{ 1+\dfrac{a^2 \Lambda}{3}}\, \varphi + \dfrac{a} 
{r_c^2+a^2} \left( 1+\dfrac{a^2 \Lambda}{3}\right) t
\end{align}
where $r_c$ represents the position of the cosmological horizon 
and is the largest of the three positive roots of the polynomial 
$\Delta_r$, the other two roots being still labelled by $r_+$ 
(the event horizon) and $r_-$ (the Cauchy horizon).\pid
The line element (1) becomes
\begin{multline} \hspace{-0.5cm}
ds^2 = \dfrac{(r_c^2+a^2 \cos^2 \vartheta )^2 \Delta_r -a^2 
\sin^2 \vartheta \left( 1+\dfrac{a^2 \Lambda }{3} \cos^2 
\vartheta \right) (r_c^2-r^2)^2}{(r_c^2+a^2)^2 (r^2+a^2 \cos^2 
\vartheta )} \\ {} \\ \hspace{-0.5cm}
-\, \dfrac{r^2+a^2 \cos^2 \vartheta}{\Delta_r}\, dr^2 -\, \frac
{r^2+a^2 \cos^2 \vartheta}{1+\dfrac{a^2 \Lambda}{3} \cos^2 
\vartheta}\, d\vartheta ^2  \\ {} \\ \hspace{-1.9cm}
+ \dfrac{2a \sin^2 \vartheta \left[
\left(1+\dfrac{a^2 \Lambda }{3} \cos^2 \vartheta
\right)(r_c^2-r^2)(r^2+a^2) -\, (r_c^2+a^2 \cos^2 \vartheta )
\Delta_r \right]} { \sqrt{1+\dfrac{a^2 \Lambda}{3}}(r_c^2+a^2)  
(r^2+a^2 \cos^2 \vartheta )} dt d\varphi \\ {} \\ 
-\, \dfrac{\sin^2 \vartheta \left[\left(1+\dfrac{a^2 \Lambda}{3}
\cos^2 \vartheta \right)(r^2+a^2)^2 - a^2 \sin^2 \vartheta \Delta_r 
\right] }{ \left( 1+\dfrac{a^2 \Lambda}{3}\right) 
(r^2+a^2 \cos^2 \vartheta)}\, d\varphi ^2 
\end{multline}
In this case the curve of the extreme black holes in the plane 
$(M^2 \Lambda /3,\, a^2/M^2)$, which is given again by Eq.($8$),
begins at the point $(0,1)$ and 
ends at the point $\left(\dfrac{16}{135+78 \sqrt{3}}, \, \dfrac 
{3+2 \sqrt{3}}{16}\right)$ where the three positive roots of the 
polynomial $\Delta_r$ have the same value equal to $\dfrac{(3+2 
\sqrt{3})M}{4}$; the corresponding plot is shown in Fig.1.\pid
The angular velocity $\omega$ is given by
\begin{equation} \hspace{-2.0cm}
\omega = \dfrac{a\, \sqrt{1+\dfrac{a^2 \Lambda}{3}}  
\left[\left(1+ \dfrac{a^2 \Lambda}{3} \cos^2 \vartheta 
\right)(r_c^2-r^2)(r^2+a^2) -(r_c^2 +\cos^2 \vartheta ) \Delta_r
\right]}{(r_c^2+a^2)\left[\left(1+ \dfrac{a^2 \Lambda}{3} \cos^2 
\vartheta \right)(r^2+a^2)^2 - a^2 \sin^2 \vartheta \Delta_r 
\right] } 
\end{equation}
and as requested goes to zero as $r \to r_c$ ; we notice that, as
expected, also $g_{tt}$  goes to zero in this limit. A plot of 
$\omega$ as a function of $r_+$: 
\begin{equation}
\omega = \dfrac{a (r_c^2-r_+^2) \sqrt{1+\dfrac{a^2 
\Lambda}{3}}}{(r_c^2+a^2)(r_+^2+a^2)}
\end{equation} 
is given in  Fig.2.\pid
The area of the horizon can again be written as
\begin{equation}
A = 4\pi \dfrac{r_+^2 + a^2}{\sqrt{1+\dfrac{a^2 \Lambda}{3}}}
\end{equation} 
The plot of $A$ as a function of $r_+$ is given in Fig.3 and 
shows that $A$ increases monotonously as $r_+$ goes from $ M$ 
to $r_c$.\pid
Some concluding remarks seem here appropriate.\pni
a) While in the Kerr metric ($\Lambda = 0$) the observer is put 
at infinity where $g_{tt} = 1$, in the case $\Lambda \neq 0$ 
all the pairs  $(r^*, \vartheta^*)$ solutions to $g_{tt} = 1$
and fixing an observer should be considered. 
If then one wants to put the observer at a predefined 
position $(r_0, \vartheta_0)$ outside the ergosphere, 
it simply suffices to make the change of variable 
\begin{equation}
t = \dfrac{\bar{t}}{\sqrt{g_{tt}(r_0,\vartheta_0)}}
\end{equation} 
which in turn modifies $\omega$ only by a scale factor.\pni
b) If we consider, when $\Lambda < 0$, the area of a surface at
constant $t$ and $r$, and the lengths of the closed curves on 
it, we see that, while our coordinate transformations on $\psi$ 
and $\chi$ give asymptotically the correct value $2\pi r \sin 
\vartheta$ for a closed azimuthal curve at polar angle 
$\vartheta$, it is not possible to recover the asymptotically 
expected values $4\pi r^2$ and $2\pi r$ respectively for the area
of a surface of radius $r$ and for the length of a polar curve 
$\varphi$ = constant. The drawback is due to the particular form
of the term $\Delta_\vartheta$ which appears in the 
$g_{\vartheta\vartheta}$ component of the metric tensor. That 
term could be eliminated by the change of variable
\begin{equation}
\bar{\vartheta} = 
\int{\dfrac{d\vartheta}{\sqrt{\Delta_\vartheta}}} 
\end{equation}
but it would then be unpossible to express analytically 
$\vartheta$ as a function of $\bar{\vartheta}$. We notice 
however that in calculating areas related to black holes, as well
as to extreme black holes as made here, the term 
$\Delta_\vartheta$ gets simplified in calculations by the use of 
the condition $\Delta_r = 0$. \pni 
c) Finally, the fact that the Kerr-dS Universe is closed 
requires the presence of another antipodal mass $M$ equal to the 
mass of the original source and endowed with equal but opposite 
angular momentum. 
\newpage
\begin{flushleft}
\Large\textbf {Figure captions}
\end{flushleft}
Figure 1: The curve of the extreme black holes. Here 
$x=M^2\Lambda /3, \; y=a^2/M^2$. A dashed 
line separates the two regions where $\Lambda$ takes opposite 
signs. \pni 
Figure 2: The angular velocity $\omega '$ as a function of 
$r_+ '$. The ordinate axis separates the regions where $\Lambda <
0$ (left) and where $\Lambda > 0$ (right). \pni
Figure 3: The area $A'$ as a function of $r_+ '$. The ordinate
axis separates the regions where $\Lambda < 0$ (left) and where 
$\Lambda > 0$ (right). \newpage
\begin{figure}
    \begin{center}
   \includegraphics[bb=0 -50 500 400]{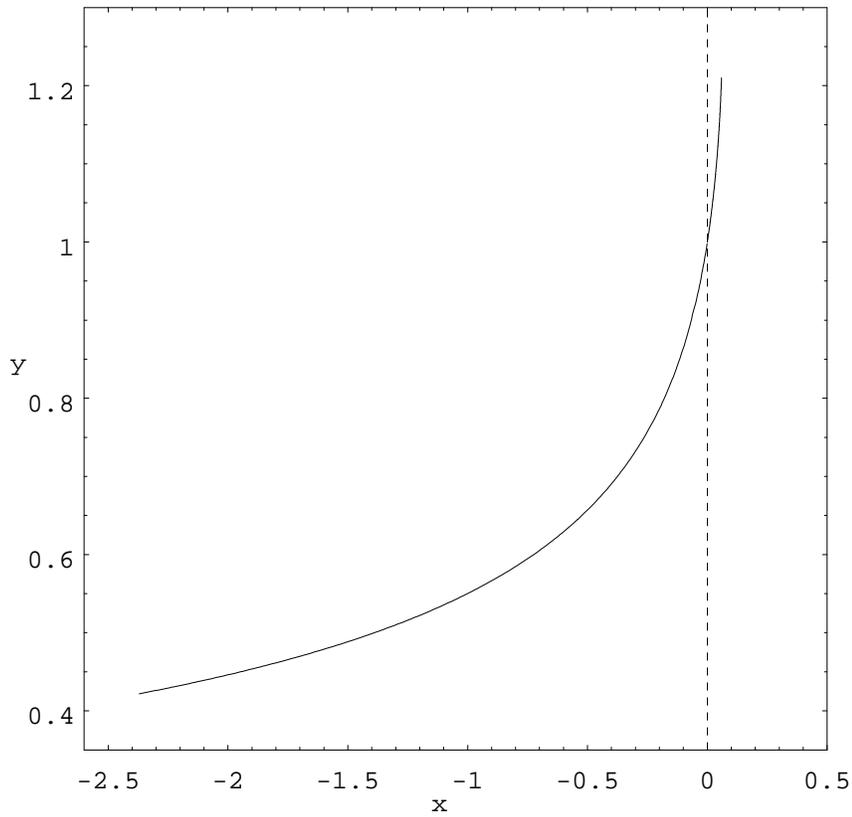}
    \caption{The curve of the extreme black holes. Here 
$x=M^2\Lambda /3, \; y=a^2/M^2$.  A dashed 
line separates the two regions where $\Lambda$ takes 
opposite signs.} 
\end{center}
\end{figure}
\newpage    
 \begin{figure}
    \centering
   \includegraphics[bb=80 -50 600 400]{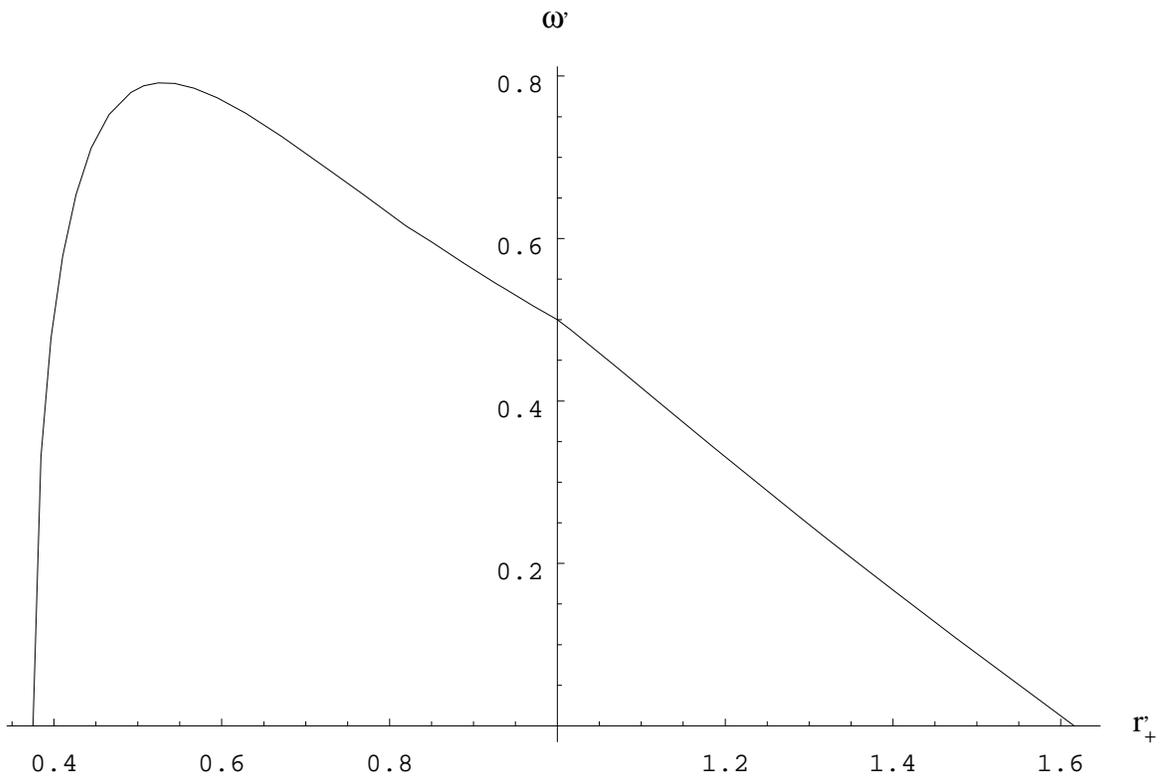}
    \caption{The angular velocity $\omega '$ as a function of 
$r_+ '$. The ordinate axis separates the regions where $\Lambda 
< 0$ (left) and where $\Lambda > 0$ (right).} 
\end{figure}
\newpage
\begin{figure}
    \centering
   \includegraphics[bb=0 -50 500 400]{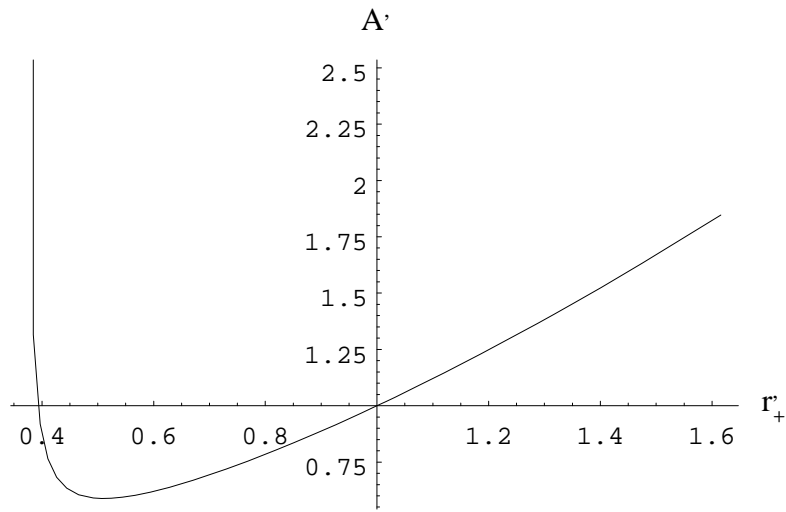}
    \caption{The area $A'$ as a function of $r_+ '$. The
    ordinate axis separates the regions where 
    $\Lambda < 0$ (left) and where $\Lambda > 0$ (right).}
\end{figure}
\end{document}